# Integration of Computational Techniques for the Modelling of Signal Transduction

Pedro Pablo González Pérez
Maura Cárdenas García
Carlos Gershenson García
Jaime Lagúnez-Otero

Instituto de Química
Universidad Nacional Autónoma de México
Ciudad Universitaria, 04510, MÉXICO, D.F.
ppgp@servidor.unam.mx http://132.248.11.4

*Abstract*: A cell can be seen as an adaptive autonomous agent or as a society of adaptive autonomous agents, where each can exhibit a particular behaviour depending on its cognitive capabilities. We present an intracellular signalling model obtained by integrating several computational techniques into an agent-based paradigm. Cellulat, the model, takes into account two essential aspects of the intracellular signalling networks: cognitive capacities and a spatial organization. Exemplifying the functionality of the system by modelling the EGFR signalling pathway, we discuss the methodology as well as the purposes of an intracellular signalling virtual laboratory, presently under development.

## 1. Introduction

Each cell in a multicellular organism receives specific combinations of chemical signals generated by other cells or from their internal milieu. The final effect of the signals received by a cell can be translated in the regulation of the cell metabolism, in cellular division or in its death. Once the extracellular signals bind to the receptors, different signalling processes are activated, generating complex information transmission networks.

The more experimental data about cellular function we obtain, the more important the computational models become. The models allow for the visualization of the network components and permit the prediction of the effects of perturbations on components or sections of the signalling pathway.

Within the computer sciences, the artificial intelligence is one of the main areas to model biological systems. This is due to the great variety of models, techniques and methods that support this research area, many of which are inherited from disciplines such as cognitive sciences and neuroscience. Among the main techniques of artificial intelligence and computer sciences commonly used to model cellular signalling networks are artificial neural networks (Bray and Lay, 1994; Pritchard and Dufton, 2000), Boolean networks (Edwards, 1995), petri nets (Fuss, 1987), rule-based systems (Cárdenas-García, 2001), cellular automata (Edwards, 1995; Wurthner, Mukhopadhyay and Peimann, 2000), and multi-agent systems (Fisher, Paton and Matsuno, 1999; Paton, Staniford and Kendall, 1995; Schwab and Pienta, 1997).

The high complexity level of intracellular communication networks makes them difficult to model with any isolated technique. However, when integrating the most relevant features of these techniques in a single computational system, it should be possible to obtain a more robust model of signal transduction. This would permit a better visualization, understanding of the processes and components that integrate the networks.

The theory of behaviour-based systems constitutes an useful approach for the modelling of intracellular signalling networks (González, Gershenson, Cárdenas and Lagúnez-Otero, 2000). The model permits to take into account communication between agents via a shared data structure, in which other cellular compartments and elements of the signalling pathways can be explicitly represented. In this sense, the blackboard architecture (Nii, 1989) becomes appropriate.

In this paper, we demonstrate that an effective and robust model of intracellular signalling can be obtained when the main structural and functional characteristics of behaviour-based systems are joined with the blackboard architecture. That is, a cell can be seen as a society of autonomous agents, where each agent communicates with the others through the creation or modification of signals on a shared data structure, named "blackboard". The autonomous agents model determinated functional components of intracellular signalling pathways, such as signalling proteins and other mechanisms. The blackboard levels represent different cellular compartments related to the signalling pathways, whereas the different objects created on the blackboard represent signal molecules, activation or inhibition signals or others elements belonging to the intracellular medium.

In this way, when the autonomous agents are used in an intracellular signalling model, the cognitive capabilities of the signalling pathway can be taken into account. On the other hand, the use of blackboard architecture permits to model the high level of spatial organization exhibited by the intracellular signalling networks and structural characteristics of the intracellular medium.

The rest of the paper is structured as follows: the next section presents an overview of the information processes of signalling intracellular networks. In section 3, some computational models for intracellular signalling are discussed; section 4 presents the intracellular signalling model proposed; section 5 describes the methodology steps for the creation of an intracellular signalling model using Cellulat. In section 6 an instantiation of the paradigm is made with the EGF receptor signalling pathway. Section 7 is focussed on the final goal of this work: the creation of an intracellular signalling virtual laboratory.

## 2. An overview of signalling intracellular networks

In multicellular organisms, cell decisions about survival, growth, gene expression, differentiation and senescence or death, are made on the basis of external signals. These stimuli include cell-cell adhesion, growth factors, hormones, cytokines, neuropeptides and ions. The skill to integrate information from multiple sources is essential for the ability of the cell to respond appropriately to a wide range of conditions, and therefore enhances the adaptability and survival of the organism.

Signal transduction networks allow cells to perceive changes in the extracellular environment in order to produce an appropriate response. One of the most exciting recent development in molecular and cell biology has been the step by step construction of signaling cascades that trace the path of the effects of an extracellular stimulus from the external membrane to the cell nucleus. A cellular process network mediates the transmission of extracellular signals to their intracellular targets.

In general, the external signals are transmitted to the interior of the cells through receptors activating diverse signaling pathways. They can follow a single way and generate an answer or a specify cellular final process, or branch out and give rise to others. These pathways considered as a whole form an interconnected network, because pathways corresponding to different stimuli cross and generate alternative trajectories.

The intracellular signalling implies several molecular processes. The signals can be as simple as the direct introduction of the signal to the nucleus and the activation of the transcription of proteins involved in the specific cellular function, which is expected. On the other hand, they can be very complicated and include multiple stages. For example, the receptor activates effector proteins like second messengers, kinases or phosphatases. They, in turn, activate transcription factor proteins, which determine the transcription of genes codifying for proteins involved in the specified cellular function.

Several mechanisms of signal transduction systems have been described for the extracellular signal membrane receptors; some of them include:

- Adenilate cyclase system
- Phosphoinositides-calcium system.
- Mitogen activated protein kinase system (MAPK).
- JAK/STAT activation system.
- The esfingomyeline-ceramide system.
- The ionic channel function receptor system.

The MAPK cascade is a central signal transduction pathway that is activated by growth factors, and is known to be involved in diverse cellular functions. In particular, we modeled the pathway activated by EGF since this pathway has crosstalk with other mentioned systems. The systems related to aging and cellular proliferation constitute our main goal.

We are not modelling a specific cellular type. Once we have the functional model with all the elements involved in one pluripotential cell, it is relatively easy to extend it to cellular types like epidermic cells or hepatocites, etc.

## 3. Computational models for intracellular signalling

Computational models in signal transduction pathways have been made using different points of view. Each research group chose the approach which seemed best for them and applied the most adequate computational tool for their purpose. The different models have been proposed according to the perspective that each research group has of the pathway they want to model. This perspective involves a range from the types of information processing present at cellular level, such as sequential, parallel, distributed, concurrent and emergent; to the cognitive capabilities exhibited by certain signal transduction pathway component, such as memory, learning, pattern recognition and handling fuzzy data.

In this sense, several computational approaches have been proposed to model the cellular signalling pathways, such as artificial neural networks (Bray and Lay, 1994; Pritchard and Dufton, 2000), Boolean networks (Edwards, 1995), petri nets (Fuss, 1987), rule-based systems (Cárdenas-García, 2001), cellular automata (Edwards, 1995; Wurthner, Mukhopadhyay and Peimann, 2000), and multi-agent systems (Fisher, Paton and Matsuno, 1999; Paton, Staniford and Kendall, 1995; Schwab and Pienta, 1997). Table 1 summarizes the main characteristics of these computational approaches, taking into account the idea behind the approach, the cognitive capabilities that can be modelled, types of present information processing, and the part of the cellular signalling to be modelled.

From each model we took some points that we considered suitable for our work. Thus, our approach sees the dynamic of the cellular signal transduction in terms of a collection of autonomous agents communicating between them through a shared data structure, where each agent is implemented as an artificial neural network, a package of production rules, a Boolean network or a molecular automata, depending of the complexity of the task carried out by the agent and the knowledge degree or cognitive capabilities required by it.

## 4. Modelling intracellular signalling pathways using the adaptive autonomous agents approach

| Computational approach | Idea behind the approach | Cognitive capabilities | Types of information processing | Cellular signalling pathway modelled | References |
|---|---|---|---|---|---|
| Boolean networks | The cell can be modelled as a network of two-sate components interacting between them. The state of each component in the network depends of a particular boolean function. | Boolean logic computation | Parallel | Intracellular signalling Genetic networks | Kauffman, 1991; Edwards, 1995; Karp and Paley, 1994; Armas et. al., 2000 |
| Expert systems | The interactions (activation, phosphorylation, etc.) between signalling network components are modelled using production rules | knowledge-based inference and deduction | Sequential Parallel | Intracellular signalling Mitogen activated protein kinase system | Lagúnez-Otero, 1998; Cárdenas, 2001; Takai-Igarashi, T., and Kaminuma,T.1998 |
| Cellular automata | The interaction between cells or molecules is modelled as a matrix, where the state of an element of the matrix depends on the states of the neighbouring elements. | None | Parallel | Extracellular signalling Intracellular signalling | Marijuan, 1994; Levy, 1992; Wurthner, Mukhopadhyay and Peimann, 2000 |
| Petri nets | The cell is seen as a connected graph with two types of nodes. One type represents elements, such as signalling molecules and proteins, whereas the other type represents transitions, such as activations. | None | Sequential Concurrent | Intracellular signalling | Holcombe, 1994 |
| Artificial neural networks (ANN) | The proteins in signalling networks are seen as artificial neurons in ANN. Like an artificial neuron, a protein receives weighted inputs, produces an output, and has an activation value. | Memory Learning Pattern recognition | Distributed Parallel Emergent | Glucagon Signalling Pathway Intracellular Signalling Phosphoinosites/Ca2+ pathway | Bray, 1990; Bray and Lay, 1994; Bray, 1995; Pritchard and Dufton, 2000; Paton, 1993 |
| Distributed systems (agents) | The cell is seen as a collection of agents working in parallel. The agents communicate between them through messages. | Memory Learning Pattern recognition Handling fuzzy data Adaptive action selection | Distributed Parallel Emergent | Intracellular signalling | Paton, Staniford and Kendall, 1995; Fisher, Paton and Matsuno, 1999 |

Table 1. Summary of different computational approaches for the modelling of cellular signalling.

## 4.1. Adaptive autonomous agents

The theory of behaviour-based systems (Brooks, 1986, Maes, 1994) provides a new philosophy for the construction of autonomous agents, inspired by ethology. The goal opportunistically to changes on the blackboard, some mechanism is necessary to control these changes and to decide, at each relevant moment, which actions should be taken. The control mechanism handles the interaction between the blackboard, the knowledge sources, and the outsourcing; such as users and control or data acquisition subsystems.

The knowledge sources can be built as adaptive autonomous agents when an adaptive action selection mechanism has been allowed to control the interactions that the first ones execute on blackboard (González, 2000). In this way, these agents can model intracellular signalling pathway components such as receptors, proteins and enzymes. On the other hand, the blackboard structure allows to model the intracellular medium structure through its levels. That is, different cellular structures involved in the intracellular signalling could be mapped to different blackboard levels. In this sense, cellular membrane, juxta-membrane region, cytoplasm and nucleus will constitute different blackboard levels. The blackboard structure also provides a continuous trace of all interactions occurred between the agents. This trace can be seen as a topologic map distributed between the blackboard levels and its elements reflect the different activation, inactivation, phosphorylation and desphosphorylation degree that characterize the intracellular signalling in a given time.

These previous considerations constitute the functional and structural essence of the intracellular signalling model described below.

### 4.3. The model

Our proposal consists in modelling the cell as an autonomous agent, which in turn is composed by a society of autonomous agents, where each agent communicates through a blackboard with others. The model proposed here constitutes a refinement and adaptation of an action selection mechanism structured on a blackboard architecture previously developed by us, called Internal Behaviours Network (IBeNet) (Gershenson, González and Negrete, 2000a; Gershenson, González and Negrete, 2000b; González, 2000; González, Negrete, Barreiro and Gershenson, 2000). Although the IBeNet was initially built to control the action selection in autonomous agents (physical robots, animats, or artificial creatures simulated on a computer), it constitutes a working environment for the bottom-up modelling of information processing systems characterized by: (1) modularity, (2) parallel, distributed and emergent processing, (3) coordination and opportunistic integration of several tasks in real time, (4) use of several abstraction or context levels for the different types of information that participate in the processing network, (5) decision making, and (6) cognitive capabilities such as adaptive action selection, memory and learning.

It is known that the information processing at a cellular level is characterized by many of the properties mentioned before. In this sense, the model proposed must constitute a good approach to model the intracellular signalling pathways.

As mentioned above, the intracellular signalling model has been named Cellulat (a kind of animat (artificial animat) which behaves as a cell). In Figure 2, the structure of Cellulat can be appreciated. Three main components define the Cellulat structure: the blackboard, the internal autonomous agents and interface autonomous agents.

The blackboard represents the cellular compartments. Different levels in the blackboard correspond to different cellular compartments through which the signal transduction take place. In this way, the cellular membrane, the cytoplasm and the nucleus could be represented as different blackboard levels. The solution elements recorded on the blackboard represent two main types of intracellular signals: signalling molecules and activation/inactivation signals. Both types of signals are synthesized or created by internal autonomous agents and these, either directly or indirectly, promote the activation/inactivation of other internal autonomous agents. Other types of cellular elements or structures can be represented on the blackboard as well.

The term "internal autonomous agents" has been used to identify the autonomous agents whose tasks deal with the creation or modification of signals on the blackboard. An internal autonomous agent obtains a signal or combination or signals from a determined blackboard level and transduces these into other signals on the same or other blackboard level. The way in which a signal is transduced depends of the cognitive capabilities of the internal autonomous agent. Internal autonomous agents model components of the intracellular signalling network such as proteins and other molecules necessary to carry out the signal transduction

On the other hand, the function of an interface autonomous agent is to establish the communication between the blackboard and the external medium (they are similar to sensors or actuators in BBS). Not all external signals or combinations of these are recognized by an interface autonomous agent; this recognition depends both of the signal characteristics and the cognitive capabilities of the interface autonomous agent. Interface autonomous agents model the cell surface receptors and the mechanisms for the production of signalling molecules.

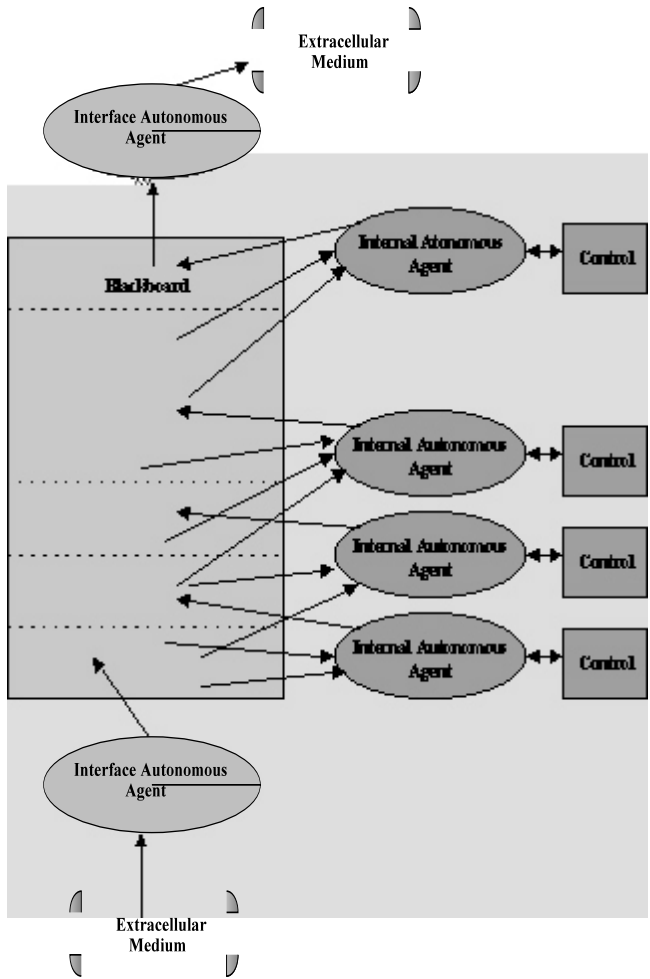

Figure 2 Architecture of the Cellulat.

Each agent, independently of its type, has a condition part and an action part. The way in which both parts are linked depends on the complexity of the intracellular component modelled by the agent. For this reason, agents which model complex components could use more advanced techniques, such as neural networks, or any combination of other techniques, to link both parts. Agents which model less complex components could use less sophisticated but useful techniques, such as production rules, Boolean networks or others. The work of both types of agents is event-directed. This is, each intracellular signal registered on the blackboard or each extracellular signal perceived constitutes an event, which could activate or inactivate one or more autonomous agents. When an internal autonomous agent is activated then it executes its action, consisting in the creation or modification of a signal on the blackboard.

In Table 2 and Table 3 the characteristics and functionality of two intracellular signalling components in Cellulat are shown. Table 2 shows an interface autonomous agent whose identity, attributes and behaviour repertoire correspond to a type I receptor, whereas Table 3 represents an internal autonomous agent which models an adapter protein. In both cases, the behaviours of these components have been modelled using production rules, as a first approach.

**Receptor**

Identity (I)
Structural state (SS)
Phosphorylation State (PS)
Activation State (AS)
Ligand vector (LV)
Phosphorylation Site Vector (PSV)
Interaction Protein Vector (IPV)
Cellular Compartment (CC)

Create-Receptor ( )
External-Interaction ( )
Phosphorylation ( )
Activation ( )
Internal-Interaction ( )
Destroy-Receptor ( )

Table 2. Receptor model in Cellulat.

**Adapter Protein**

Identity (I)
Domain Vector (DV)
Interaction Protein Vector (IPV)
Cellular Compartment (CC)
Initial Inactive Concentration (IIC)
Actual Active Concentration (AAC)
Inactive Actual Concentration (IAC)

Create-Protein( )
Protein-Receptor-Interaction ( )
Protein-Protein-Interaction ( )
Change-State-Domain( )

Table 3. Adapter protein model in Cellulat.

### 4.4. Cognitive capabilities of Cellulat's agents

It is known that certain types of proteins exhibit several cognitive capabilities such as pattern recognition, memory, and handling fuzzy data (Fisher, Paton and Matsuno, 1999). The evolution, learning and emergence of properties have also been suggested in protein networks (Pritchard and Dufton, 2000). In Cellulat, both types of autonomous agents exhibit several cognitive capabilities including pattern recognition, handling uncertainty, and fuzzy data, adaptive action selection, memory, and learning. These capabilities allow the autonomous agents to exhibit adaptive behaviour. These cognitive capabilities are supported by several artificial intelligence paradigms including the following approaches: rule-based reasoning, probabilistic reasoning, artificial neural networks, Boolean networks, fuzzy logic systems, and multidimensional logic systems.

### 4.5. Modelling spatial organization in Cellulat

Another important aspect to consider in the modelling of intracellular signalling networks is their spatial organization. Recent experimental data clearly demonstrate that many intracellular signalling networks exhibit a high level of spatial organization. Cellular functions could depend on the spatial organization of the cell's components. New intracellular signalling models take this into account (Fisher, Malcom and Paton, 2000). Cellulat allows to model the spatial organization taking into account two organizational criteria of the blackboard architecture. The first is horizontal organization, given by the different abstraction levels of the blackboard, which allow intra level signal processing. The other is vertical organization, given by columns that vertically cross different blackboard levels. These columns arise as result of the adjoining work of several internal autonomous agents that operate at a same section of blackboard, which cover different blackboard levels. We have named these columns "agency columns". These columns represent signalling pathways.

Convergence and divergence of agency columns could occur, and these processes could be related with evolution and learning of the signalling networks (González, 2000). In this way, the model proposed allows to model intracellular signalling pathways by taking into account their spatial organization. This is, the two information processing levels present in Cellulat (horizontal and vertical) allow us to establish what may be called a "topology preserving map".

## 5. Methodology

Once Cellulat has been implemented, it constitutes a type of shell to model particular signal transduction pathways. The creation of an intracellular signalling network can be seen as an incremental process of definition and refinement of the pathway elements such as receptors, enzymes and secondary messengers, which are expressed as interface autonomous agents, internal autonomous agents or objects recorded on blackboard levels in Cellulat. In this sense, the creation of an intracellular signalling network involves the application of the following methods:

1. Select an individual signalling pathway to model.
   a. Divide the signalling pathway in sections to model and test. We have consider the following three sections:
      i. First Section: from ligand to the activation of the receptor.
      ii. Second Section: from the activated receptor to the activation of cytosolic kinases.
      iii. Third Section: from the activated cytosolic kinases to the activation of transcription factors.
   b. Identify and define the components belonging to the pathway section. That is, internal autonomous agents such as proteins, interface autonomous agents such as receptors, and objects recorded on blackboard levels such as activation/ phosphorylation signals and signalling molecules.
   c. Test the signalling pathway section, displaying the resultant behaviour through activity-time and concentration-time curves and activity maps.
   d. If the resultant behaviour of the model is not desirable, adjust and refine the signalling pathway section.
   e. If there are still signalling pathway sections to model then go to step 1.b, otherwise go to next step.
2. Join the different signalling pathway sections to form the complete pathway and to test this one.
   a. Adjust and refine the description of the individual signalling pathway, displaying the resultant behaviour through activity-time and concentration-time curves and activity maps.
   b. If there is another intracellular signalling pathway to model then go to step 1. Else, go to

next step.

3. Define connections of two or more individual pathways, taking into account the following criterion:
- Second messengers produced by one pathway are used as inputs to other pathways.
- Enzymes whose activation was regulated by one pathway are connected to substrates belonging to other pathways.
   a. Test the signalling network behaviour against published experimental data.
   b. If the network resultant behaviour is not expected then refine the connections between pathways and return to step 3.a. Else, go to next step.
4. The model of the signalling network has been completed.

The definition of the signalling pathway components consists in the creation of Cellulat's agents. This process can be seen as the creation of new instances from templates declared and contained in Cellulat. A new instance is created defining the appropriate values for the attributes specified as part of an instance state. On the other hand, the adjustment and refinement of an individual signalling pathway, signalling pathway sections or interaction between pathways is reduced to the modification of different attribute values belonging to certain agents in Cellulat.

## 6. Modelling EGF receptor signalling pathway: an initial approach

Applying part of the methodology previously presented we modelled a MAPK signalling pathway activated by the epidermal growth factor (EGF). We began with this pathway because some of its elements participate in other pathways that we will model afterwards. Figure 3 shows an interaction model of the MAPK signalling pathway. The identity of proteins belonging to MAPK signalling pathway are shown in Table 4. In this section we will present only a few steps of the previously described methodology.

The EGF receptor (EGFR) signal pathway is one of the main regulators of cell proliferation, differentiation, senescence and apoptosis. For the modelling of this pathway we divided it in three parts. The elements included in these parts are explained below.

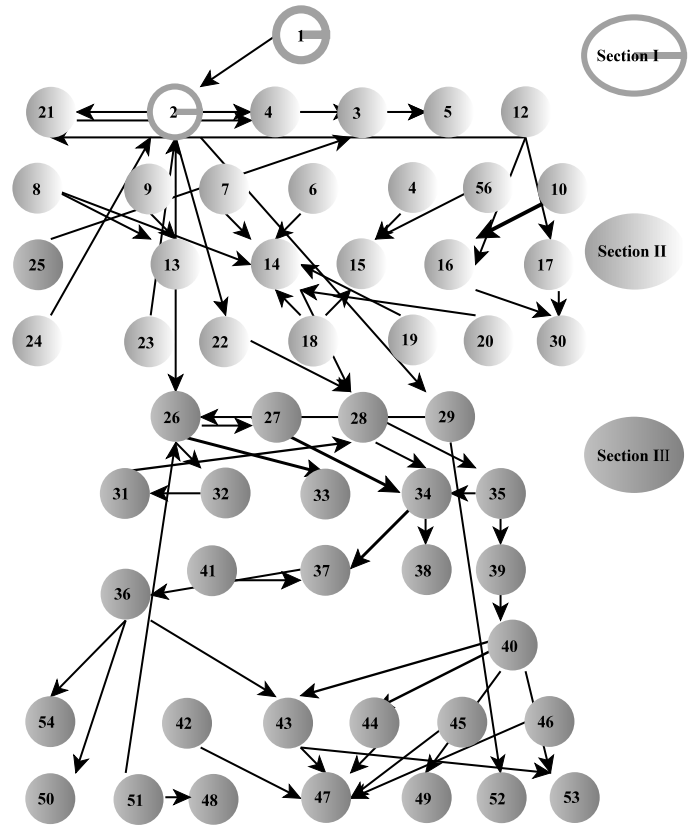

Figure 3. MAPK singalling pathway interaction model. Numbers are individual proteins. Arrows represent relation between two proteins. Two or more arrows mean cooperative or alternative action. Sections are indicated.

| 1 EGF | 7 p120 | 13 Rho | 19 RasGRp | 25 Pyk2 | 31 PKC | 37 ERK1 | 43 cJun | 49 Elk1 |
|---|---|---|---|---|---|---|---|---|
| 2 EGFR | 8 p190 | 14 Ras | 20 RasGRF | 26 PI3K | 32 PDK | 38 ERK2 | 44 JunD | 50 CREB |
| 3 Grb2 | 9 p115 | 15 Rap | 21 Src | 27 PAK | 33 AKT | 39 JNKK | 45 FosB | 51 cAbl |
| 4 Shc | 10 cdc42 GAP | 16 cdc42 | 22 PLCg | 28 Raf | 34 MEK | 40 JNK | 46 JunB | 52 STAT |
| 5 Gab2 | 11 RapGAP | 17 Rac | 23 SHP2 | 29 JAK | 35 MEKK | 41 MKP1 | 47 AP1 | 53 ATF2 |
| 6 Sos | 12 Dbl | 18 KSR | 24 SHP1 | 30 MLK | 36 RSK | 42 cFos | 48 IP1 | 54 SRF |

Table 4. MAPK signal transduction elements. The numbers correspond to Figure 3.

Polypeptides such as growth factors, differentiation factors and hormones are crucial components of the regulatory systems that coordinate development of multicellular organisms. Many of these factors mediate their pleiotropic actions by binding to and activating the cell surface receptors with an intrinsic protein tyrosine kinase activity. It appears that ligand-induced activation of the kinase domain and its signalling potential are mediated by receptor dimerization. Ligand binding and the subsequent conformational alteration of the extracellular domain induce receptor dimerization, which stabilizes interaction of kinase function by molecular interaction. Receptor dimerization permits the transmission of a conformational change from the extracellular domain to the cytoplasmic domain and it allows its interaction with the following proteins of the signalling cascade. These interactions give rise to the first section of the model.

The identified components belonging to this section are the following:
- **Signalling molecules**: EGF
- **Receptors**: EGFR

Table 5 presents the value assignment to the attributes defining one of the previously identified components. The elements of the Ligand Vector represents the ligand identity and affinity of the receptor for the ligand. The elements of the Phosphorylation Site Vector represent the identity of the molecule, identity of the phosphorylation protein, and the site state. When this identity is equal to the receptor identity (in this case, EGFR), it is a autophosphorylation site. The elements of the Interaction Protein Vector represent the protein identity, the domain identity and the affinity of the receptor for the protein. An affinity value equals to 1 corresponds to the greater affinity value.

| Epidermal Growth Factor Receptor (EGFR) ||
| --- | --- |
| Attribute | Value |
| Structural state (SS) | 0 /*monomer*/ |
| Phosphorylation State (PS) | 0 /*non phosphorylated*/ |
| Activation State (AS) | 0 /*inactivated*/ |
| Ligand vector (LV) | ((EGF, 2)) |
| Phosphorylation Site Vector (PSV) | ((Y920,Src,1), (Y891,Src,2), (Y1173,EGFR,1), (Y1148,EGFR,2), (Y1086,EGFR,2), (Y1068,EGFR,2), (Y992,EGFR,2), (T654,MAPK,2), (T669,MAPK,1)) |
| Interaction Protein Vector (IPV) | ((Grb2, SH2,3), (Shc, SH2,3), (PLCg, SH2,1), (PI3K, p85SH2,2)) |
| Cellular Compartment (CC) | membrane |

Table 5. Initial values of an interface agent EGFR.

Controlling the state of phosphorylation is an important mechanism by which signalling molecules regulate the activity of other proteins. A common theme in molecular signalling is the kinase cascade, in which a linear series of kinases is activated by phosphorylation by a kinase. The first kinase is specifying growth factor receptor. In this phase of the signalling cascade, phosphatases that regulates the state of activation and inactivation, and adapter proteins that facilitate interaction between two proteins also participate. This phase gives rise to the second section of the model.

The identified components belonging to the second section are the following:
- **Adapter proteins**: GRB2, SHC, Gab2
- **Guanine interchanging proteins**: Sos, p120, p190, p115, cdc42GAP, RapGAP
- **G proteins**: Rho, Ras, Rap, cdc42, Rac
- **Enzymes**: KSR, RasGRP, RasGRF, Src, PLCg, SHP-2

Table 6 presents the value assignment to the attributes defining a previously identified component in second section. The elements of the Interaction Protein Vector represent the protein identity, the domain identity and the affinity.

| GRB2 ||
| --- | --- |
| Attribute | Value |
| Domain Vector (DV) | ((SH2, free), (SH3A, free), (SH3B, free)) |
| Interaction Protein Vector (IPV) | ((Sos,SH3,1), (Gab2,SH3,2)) |
| Cellular Compartment (CC) | cytoplasm |
| Initial Inactive Concentration (IIC) | 40 nM |
| Actual Active Concentration (AAC) | 0 nM |
| Inactive Actual Concentration (IAC) | 40 nM |

Table 6. Initial values of an internal agent GRB2.

The third section of the model corresponds to the activation of transcription factors by kinase proteins. The transcription factors activate the transcription of genes codifying for proteins involved, in this case, in cell growth. Some of these formed proteins belong to this signalling pathway, and others give rise to transcription factors. This is a reason why a positive or negative feedback can be observed.

This third section includes the following components:

- **Enzymes**: Pyk2, PI3K, PAK, Raf, JAK, PAK, MLK, PKA, PKC, PDK, AKT, MEK, MEKK, RSK, ERK1, ERK2, JNKK, JNK, SEK, S6K, MKP1
- **Transcription factors**: c-Fos, c-Jun, JunD, FosB, JunB, IP-1, Elk-1, CREB, c-Abl, STAT, ATF2

In Table 7 the initial values assigned to a component in this last section are presented.

| JNKK | |
|---|---|
| Attribute | Value |
| Interaction Protein Vector (IPV) | ((JNK,Y17,1)) |
| Cellular Compartment (CC) | cytoplasm |
| Initial Inactive Concentration (IIC) | 1nM |
| Actual Active Concentration (AAC) | 0 nM |
| Inactive Actual Concentration (IAC) | 1 nM |
| Phosphorylation State (PS) | 0 /*non phosphorylated */ |
| Phosphorylation Site Vector (PSV) | ((TX, MEKK,1 )) |

Table 7. Initial values of an internal agent JNKK.

All the components of the signalling pathway MAPK identified in each of the three sections are shown in Table 8.

| Section I | |
|---|---|
| Signalling molecule | EGF |
| Receptor | EGFR |
| **Section II** | |
| Adapter protein | Grb2, Shc, Gab2 |
| Guanine interchanging protein | Sos, p120, p190, p115, cdc42GAP, RapGAP, Dbl |
| G Protein | Rho, Ras, Rap, Cdc42, Rac |
| Enzymes | KSR, RasGRP, RasGRF, Src, PLCg, SHP2, SHP1 |
| **Section III** | |
| Enzymes | Pyk2, PI3K, PAK, Raf, JAK, MLK, PKA, PKC, PDK, AKT, MEK, MEKK, RSK, ERK1, ERK2, JNKK, JNK, MKP1 |
| Trascription Factors | cFos, cJun, JunD, FosB, JunB, AP1, IP1, Elk1, CREB, cAbl, STAT, ATF2 |

Table 8. Identified components of the MAPK signalling pathway.

Figure 4 shows the semantics assigned to the different blackboard levels and types of agent in Cellulat for this pathway. For example, the agents of the "Adapter protein" type would be GRB2, SHC, and Gab2; and some of the agents of the "G-protein" type would be Rho, Ras, and Rap.

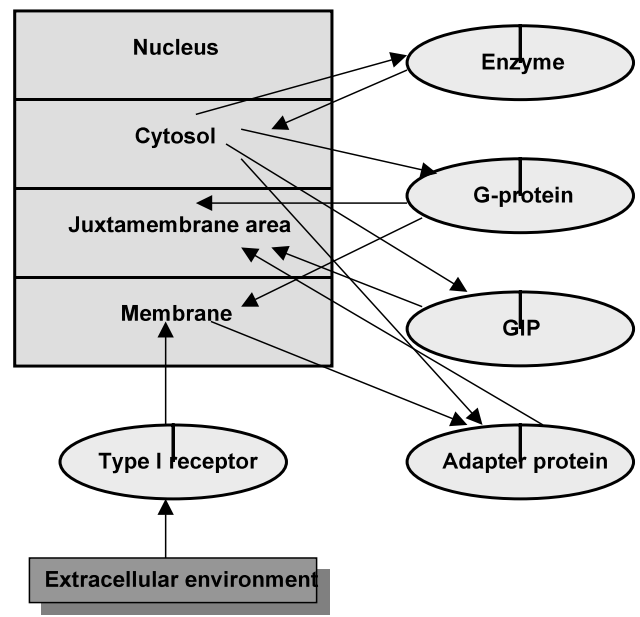

Figure 4. Model of EGFR signalling pathway in Cellulat.

Once defined the elements belonging to each of the sections of the MAPK signalling pathway, according to our methodology, each section is tested, for then assembling the sections to form the complete pathway. The individual pathway is also tested.

## 7. Towards an intracellular signalling virtual laboratory

Virtual laboratories have been developed in different areas to reproduce experiments, that in most cases, were made in real laboratories. They are used for understanding the systems they simulate, and they can also be used for teaching. Virtual labs do not replace physical labs, but they have advantages over them. They have relatively low costs, there are no inconveniences in failing experiments, and the most important thing: the researchers can control and reproduce easily situations. We will not find new structures in virtual labs, because all structures were programmed. But they are quite useful for understanding the processes and emergent properties of the systems they simulate. Thus, people can go back to the physical world, and control a system after understanding its mechanisms in a virtual lab. Virtual labs have been developed for different areas, such as physics, electronics, robotics, physiology, chemistry, engineering, economics, ecology, and ethology (Gershenson, González and Negrete, 2000).

One of the reasons for our interest in the analysis and understanding of the signalling pathways, is the possibility of regulating them. In principle, it is possible to observe this process in a virtual laboratory based on our paradigm. In particular, we would like to see the effects of perturbations on the systems such as adding elements or taking them out as knock-outs. The expected effects would be directly on the cognitive capacity of the network and ultimately on the decisions taken by the cell in order to differentiate, proliferate or become senescent. Pathologies and natural processes can be followed in the computation of the interactions made by the components in the modelled network. The paradigm presented here is the backbone of the virtual laboratory. With the virtual laboratory we hope that etiologies and expected results of putative therapeutic strategies can be visualized.

## 8. Conclusion

We have constructed an agent-based system where cognitive capabilities are coded using behaviour-based paradigms and the blackboard architecture, combined with other artificial intelligence techniques. Recruiting these techniques, the complexity of the topology and cognitive capacities of intracellular signalling system can be studied. In our group we are currently following them in the context of pathological states in particular, senescence and cancer.

## 9. Acknowledgements

We would like to thank E. St. James for excellent discussions and proofreading. Part of this work was supported by DGAPA/UNAM.